\documentclass[11pt]{article}
\usepackage{blois,epsfig}
\usepackage[latin1]{inputenc}
\usepackage[OT1]{fontenc}
\usepackage{url}

\def\be{\begin{equation}}
\def\ee{\end{equation}}
\def\bea{\begin{eqnarray}}
\def\eea{\end{eqnarray}}

\begin{document}
\vspace*{4cm}
\title{Latest Results from HiRes}

\author{D.R. Bergman, {\em for the HiRes Collaboration}}

\address{Dept. of Phys. and Astro., Rutgers Univ., 136 Frelinghuysen
  Rd., Piscataway NJ 08854, USA}

\maketitle
\abstracts{HiRes has made the first statistically significant
  observation of the Griessen-Zatsepin-Kuzmin suppression, by fitting
  the ultra high energy cosmic ray spectrum observed in monocular mode
  by the two HiRes detectors to a broken power law model.  We find the
  break to be at an energy of $5.6\times10^{19}$ eV, with a
  significance of $5.3\sigma$.  The significance is determined by
  Poisson statistics where we expect 43.2 events above the break
  point, but observe only 13.  We have also looked for correlations
  between HiRes stereo events and active galactic nuclei.  We observe
  no statistically significant correlation in a number of different
  tests.  We have performed a search for upward going showers, which
  would be indicative of neutrinos propagating through the Earth and
  interacting just below the surface.  Observing no such events, we
  set a limit on the electron neutrino flux.  This limit is
  significantly lower in this topology than limits on the other types
  of neutrinos due to the LPM effect greatly increasing the available
  target mass for neutrino interactions.  Finally we look forward to
  improving these and other results in our work in the Telescope
  Array.}

The High Resoltuion Fly's Eye Experiment (HiRes) was operated for nine
years, from June 1997 to April 2006, on Dugway Proving Grounds in
Utah, USA.  It was designed to observe ultra high energy cosmic rays
(UHECR's) using the fluorescence technique, with two sets of
telescopes placed on dessert hills separated by 12.6 km.  Details of
the fluorescence method and the analysis methods used by HiRes can be
found
elsewhere\cite{Baltrusaitis-1985-NIMA-240-410,AbuZayyad-1999-ICRC-26-5-349,Boyer-2002-NIMA-482-457}.

\section{First Observation of the GZK Suppression}

In 1966, Greisen \cite{Greisen-1966-PRL-16-748}, and Zatsepin and
Kuzmin \cite{Zatsepin-1966-JETPL-4-78}, proposed an upper energy limit
to the cosmic-ray energy spectrum. Their predictions were based on the
assumption of a proton dominated extra-galactic cosmic-ray flux which
would interact with the photons in the cosmic microwave background
(CMB) via photo-pion production.  From the temperature of the CMB and
the mass and width of the $\Delta^+$ resonance, a ``GZK'' threshold of
$\sim{6}\times{10}^{19}$~eV was calculated, and a suppression in the
cosmic-ray flux beyond this energy was predicted.  This is a strong
energy-loss mechanism that limits the range of cosmic protons above
this threshold to less than $\sim50$ Mpc.  Forty years after its
initial prediction, the HiRes experiment has observed the GZK
cutoff. In this section we describe our measurement of the flux of
cosmic rays, the resulting cosmic-ray spectrum, our analysis of this
spectrum to infer the existence of the cutoff, and our estimate of
systematic uncertainties.

HiRes data analysis is carried out in two ways. In monocular mode,
events from each detector site are selected and reconstructed
independently.  The combined monocular dataset has the best
statistical power and covers the widest energy range.  The dataset
consisting of events seen by both detectors, data analyzed in stereo
mode, has the best resolution, but covers a narrower energy range and
has less statistics.  This section presents the monocular spectra from
our two detectors.

The details of this anaysis have been now been
published\cite{Abbasi-2004-PRL-92-151101} and I will not repeat all
the details here.  The event reconstruction procedure begins with the
determination of the shower axis.  A shower-detector plane (SDP) is
determined from the pointing direction of triggered PMTs.  For the
HiRes-II monocular dataset, the PMT times are then used to find the
distance to the shower and the angle, $\psi$, of the shower within the
SDP.  This timing fit measures $\psi$ to an accuracy of
$\sim5^{\circ}$.

The number of shower particles as a function of atmospheric depth (the
amount of air traversed) is then determined.  This calculation uses
the fluorescence yield and corrects for atmospheric attenuation.  We
fit this shower profile to the Gaisser-Hillas
function\cite{Gaisser-Hillas-1977-ICRC-15-8-353}, after having
subtracting scattered \v{C}erenkov light.  This profile fit yields
both the energy of the shower and the depth at the shower maximum,
$X_{\rm max}$.

The measurement of the cosmic-ray flux requires a reliable
determination of the detector aperture.  The aperture of the HiRes
detectors has been calculated using a full Monte Carlo (MC)
simulation.  The MC includes simulation of shower development (using
CORSIKA), fluorescence and \v{C}erenkov light production, transmission
of light through the atmosphere to the detector, collection of light
by the mirrors, and the response of the PMTs, electronics and trigger
systems.  Simulated events are recorded in the same format as real
data and processed in an identical fashion.  To minimize biases from
resolution effects, MC event sets are generated using the published
measurements of the energy spectrum \cite{Bird-1993-PRL-71-3401} and
composition
\cite{AbuZayyad-2000-PRL-19-4276,AbuZayyad-2001-ApJ-557-686,Abbasi-2005-ApJ-622-910}.

\begin{figure}
  \begin{center}
    \includegraphics[width=\columnwidth]{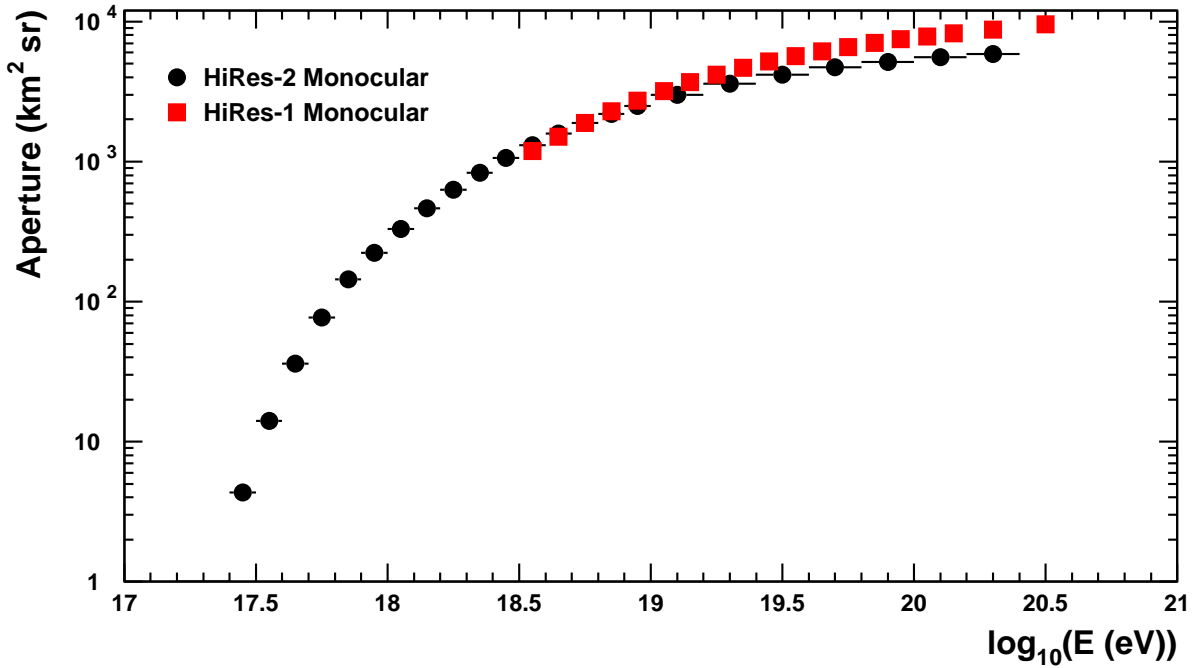}
  \end{center}
  \caption{The apertures of the HiRes-I and HiRes-II detectors
    operating in monocular mode.}
  \label{fig:aperture}
\end{figure}

To ensure the reliability of the aperture calculation, the MC
simulation is validated by comparing key distributions from the
analysis of MC events to those from the actual data.  Several of these
comparisons were shown in
reference\cite{Bergman-2007-NPBps-165-19}. Two comparisons are
especially noteworthy.  The data-MC comparison of the distances to
showers shows that the simulation accurately models the coverage of
the detector.  The comparison of event brightness shows that the
simulations of the optical characteristics of the detector, and of the
trigger and atmospheric conditions, accurately reproduce the data
collection environment.  The excellent agreement between the observed
and simulated distributions shown in these cases is typical of MC-data
comparisons of other kinematic and physical quantities, and
demonstrate that we have a reliable MC simulation program and aperture
calculation.  Figure \ref{fig:aperture} shows the result of the
aperture calculation for both HiRes-I and HiRes-II in monocular mode.

\begin{figure}
  \begin{center}
    \includegraphics[width=\columnwidth]{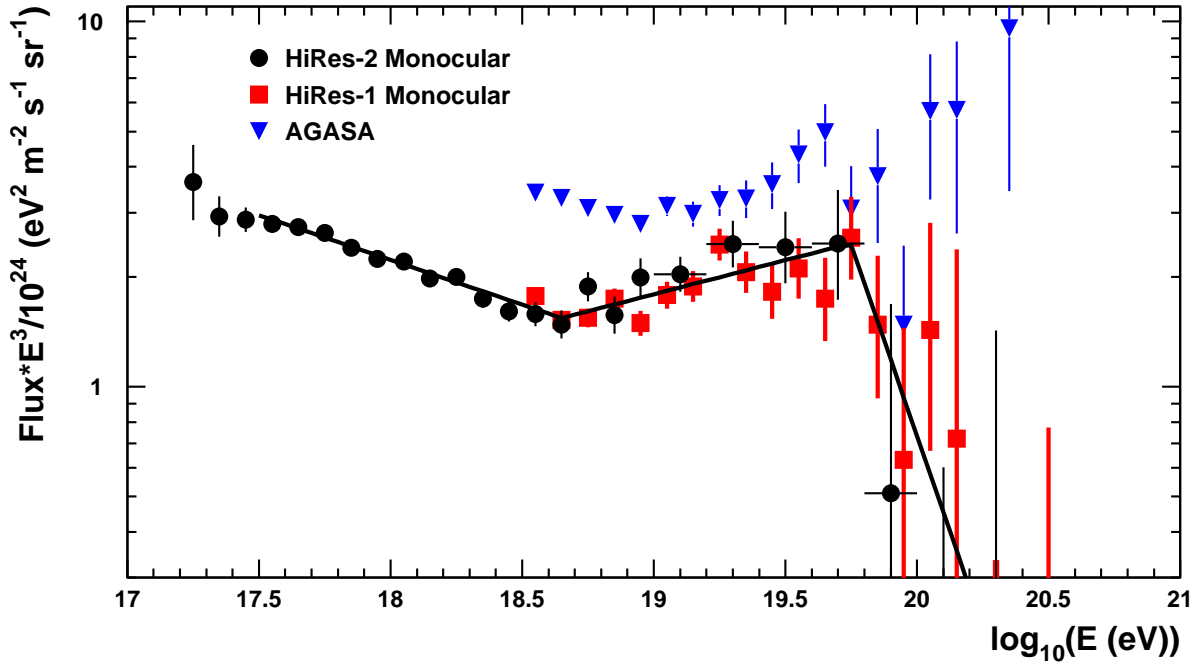}
  \end{center}
  \caption{The cosmic-ray energy spectrum measured by the HiRes
    detectors operating in monocular mode.  The spectrum of the
    HiRes-I and HiRes-II detectors are shown.  The highest two energy
    bins for each detector are empty, with the 68\% confidence level
    bounds shown.  The spectrum of the AGASA experiment is also
    shown.}
  \label{fig:spectrum}
\end{figure}

Figure \ref{fig:spectrum} shows the monocular energy spectra from the
two HiRes detectors \cite{Bergman-2007-WebSpectra}.  The data included
in the figure were collected by HiRes-I from May, 1997 to June, 2005,
and by HiRes-II from December, 1999 to August, 2004.  Figure
\ref{fig:spectrum} shows the flux multiplied by $E^3$, which does not
change the statistical interpretation of the results.  Two prominent
features seen in the figure are a softening of the spectrum at the
expected energy of the GZK threshold of $10^{19.8}$~eV, and the dip at
$10^{18.6}$ eV, known as the ``ankle''. Theoretical fits to the
spectrum \cite{Berezinsky-2006-PRD-74-043005} show that the ankle is
likely caused by $e^+e^-$ pair production in the same interactions
between CMB photons and cosmic-ray protons where pion production
produces the GZK cutoff.  The observation of both features is
consistent with the published HiRes results of a predominantly light
composition above $10^{18}$~eV \cite{Abbasi-2005-ApJ-622-910}.

At lower energies, the cosmic-ray spectrum is well fit by a piece-wise
power law model. A similar fit also gives an excellent representation
of the spectrum in Figure~\ref{fig:spectrum}. The three straight line
segments shown represent the result of a fit of the measured flux to a
triple-power law. The model contains six free parameters: one
normalization, the energies of two floating break points, and three
power law indices.

We performed a binned maximum likelihood fit
\cite{Yao-2006-JPG-33-302} to the data from the two detectors.  The
fits include two empty bins for each monocular dataset. We found the
two breaks at $\log E$ ($E$ in eV) of $19.75 \pm 0.04$, and $18.65 \pm
0.05$, corresponding to the GZK cutoff and the ankle, respectively.
When the datasets were made statistically independent by removing
events seen by both detectors from the HiRes-I dataset, we obtained a
$\chi^2$ of 35.1 in this fit for 35 degrees of freedom (DOF).  In
contrast, a fit to a model with only one break point, while able to
locate the ankle, yielded a $\chi^2$/DOF=63.0/37.  The $\chi^2$
difference of 27.9, while adding two DOF, implies that the two break
point fit is preferred at a confidence level corresponding to
$4.9\sigma$.

Another measure of the significance of the break in the spectral index
at $10^{19.8}$ eV is made by comparing the actual number of events
observed above the break to the expected number for an unbroken
spectrum. For the latter, we assume the power law of the middle
segment to continue beyond the threshold. Folding the exposures with
the overlap between the detectors removed, we expect 43.2 events above
$10^{19.8}$ eV from the extrapolation, whereas 13 events were actually
found in the data.  The Poisson probability for the observed deficit
is $\sim7.2\times10^{-8}$, which corresponds to a significance of
$5.3\sigma$.  Thus we conclude that there is a definite break in the
UHE cosmic-ray energy spectrum at an energy of $(5.6 \pm 0.5) \times
10^{19}$ eV.  Since the break occurs at the expected energy of the GZK
cutoff, we conclude that it is the GZK cutoff.

\begin{figure}
  \begin{center}
    \includegraphics[width=\columnwidth]{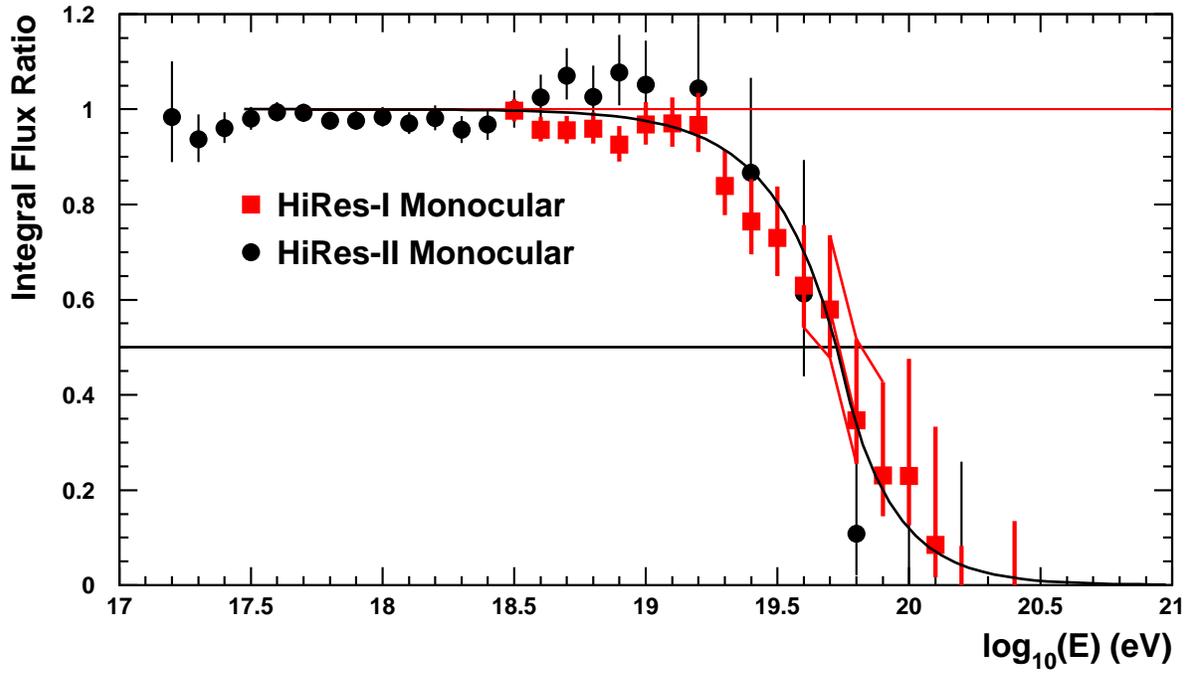}
  \end{center}
  \caption{The HiRes monocular integral spectra, divided by the
    expectation from the fit in Figure~\ref{fig:spectrum} with no high
    energy break point.  The integral spectrum from the actual fit is
    also displayed as the black line.  Only HiRes-I values (in red)
    are used to make an estimate of $E_{1/2}$, interpolating between
    the central value and one standard deviation limits.}
  \label{fig:integral}
\end{figure}

A test of this interpretation is provided by the $E_{1/2}$ method
suggested by Berezinsky and Grigorieva
\cite{Berezinsky-Grigoreva-1988-AA-199-1}.  $E_{1/2}$ refers to the
energy at which the integral spectrum falls to half of what would be
expected in the absence of the GZK cutoff.  Figure~\ref{fig:integral}
shows the integral HiRes spectra divided by the integral of the power
law spectrum used above to estimate the number of expected events
above the break.  From this plot, we find $E_{1/2} =
10^{19.73\pm0.07}$.  Berezinsky and Grigorieva predict a robust
theoretical value for $E_{1/2}$ of $10^{19.72}$ eV for a wide range of
spectral slopes \cite{Berezinsky-Grigoreva-1988-AA-199-1}. These two
values are clearly in excellent agreement, supporting our
interpretation of the break as the GZK cutoff.

We measure the index of the power law between the ankle and the GZK
cutoff to be $2.81\pm0.03$.  Above the GZK cutoff the fall-off is very
steep: we measure a power law index of $5.1\pm0.7$.  This may have
implications for the local density of extragalactic cosmic-ray sources
\cite{Berezinsky-2006-PRD-74-043005}.

For the monocular analyses, the main contributions to the systematic
uncertainty in the energy scale and flux measurements are: PMT
calibration (10\%), fluorescence yield (6\%), missing energy
correction (5\%), aerosol component of the atmospheric attenuation
correction (5\%), and mean energy loss rate ($dE/dx$) estimate (10\%).
These give a total energy scale uncertainty of 17\%, and a systematic
uncertainty in the flux of 30\%.

\begin{figure}
  \begin{center}
    \includegraphics[width=0.6\columnwidth]{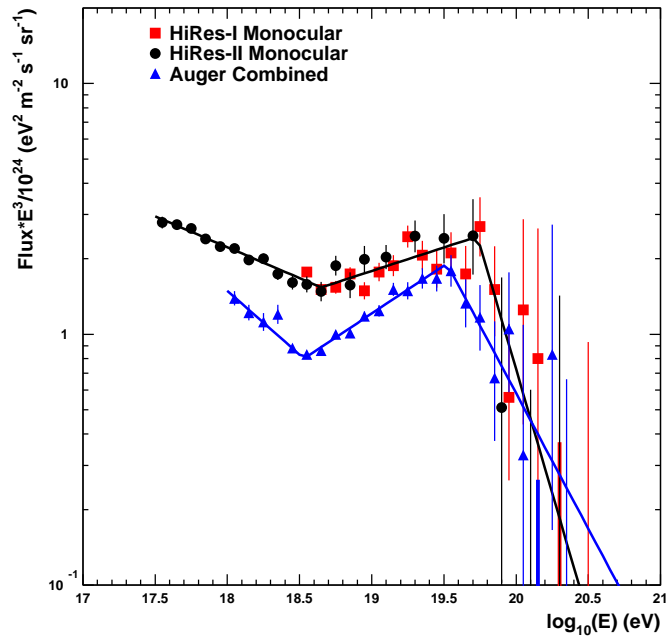}
  \end{center}
  \caption{The HiRes and Auger spectra together for comparison.}
  \label{hires-auger-spectra}
\end{figure}

The Pierre Auger Collaboration (Auger) has also recently released a
measurement of the UHECR energy
spectrum\cite{Yamamoto-2007-arXiv-0707-2638}.  This is shown in
comparison with the HiRes secptrum in
Figure~\ref{hires-auger-spectra}.  While Auger confirms the existence
of the Ankle and the GZK cutoff, there are still significant
differences.  The simplest is the over all flux level, which can be
accomodated by an approximately 10\% shift (at the GZK energy) in the
energy scal of either experiment.  More problematically, the spectral
slopes at all energies are different.  This is especially surprising
as most previous experiments agreed is their spectral slope
measurements\cite{Bergman-Belz-2007-JPG-34-R359}.

\section{Search for Correlations with AGN}

The Auger Collaboration has recently released a sky map of their
highest energy events\cite{AugerC-2007-Science-318-939}.  They report
that these events are strongly correlated with active galactic nucleii
(AGN) when they use the selection paramters $E>56$ EeV,
$\Delta\theta<3.1^\circ$ and $z_{AGN}<0.018$.  The AGN are taken from
the Veron catalog\cite{Veron-2006-AA-455-773}.  These selection
parameters were found to give the largest correlation and were the
result of a scan in the parameter space.

Since sources in the northern hemisphere may be different than in the
south, and we also see a differently shaped energy spectrum, we tested
this correlation with HiRes stereo data.  Since the HiRes flux is
larger than that seen in Auger, all HiRes energies were lowered by
10\% (see above).  These results have now been submitted for
publication\cite{Abbasi-2008-arXiv-0804-0382}.

\begin{figure}
  \begin{center}
    \includegraphics[width=0.9\columnwidth]{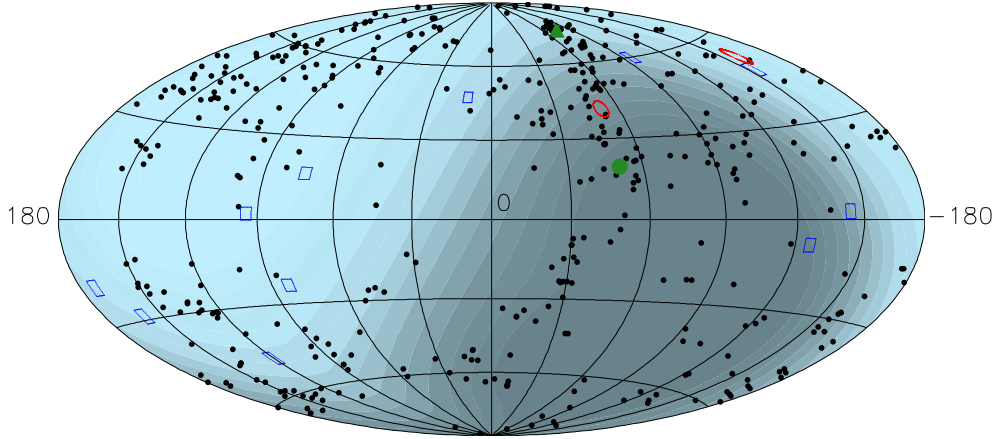}
  \end{center}
  \caption{A sky map of HiRes events above 56 EeV (blue squares and
    red boxes) and AGN with $z<0.018$ (black dots).  Correlated HiRes
    events are shown as red circles of radius $3.1^\circ$ while
    uncorrelated events are shown as blue squares.  The HiRes relative
    exposure is shown by the shading with the lightest shades being
    fully exposed and the darkes shade having no exposure.  There are
    11 shading bands for no exposure and then each 10\% of full
    exposure.}
  \label{corr-auger-point}
\end{figure}

We first looked for correlations using the Aguer paramters.  There are
13 events above 56 EeV (adjusted HiRes energy scale), but only 2 were
correlated within $3.1^\circ$ of AGN closer than $z=0.018$.  Given 13
events and the number of AGN, we expect 3.2 events to be corellated
giving a random probability of 2 or more correlation of $P_{\rm
  rand}=0.83$.  Thus, we don't confirm the Auger correlation with
their exact parameters.  The sky map of HiRes events above 56 EeV, and
AGN used as sources ($z<0.018$) along with the HiRes exposure is shown
in Figure~\ref{corr-auger-point}.

\begin{figure}
  \begin{center}
    \includegraphics[width=0.9\columnwidth]{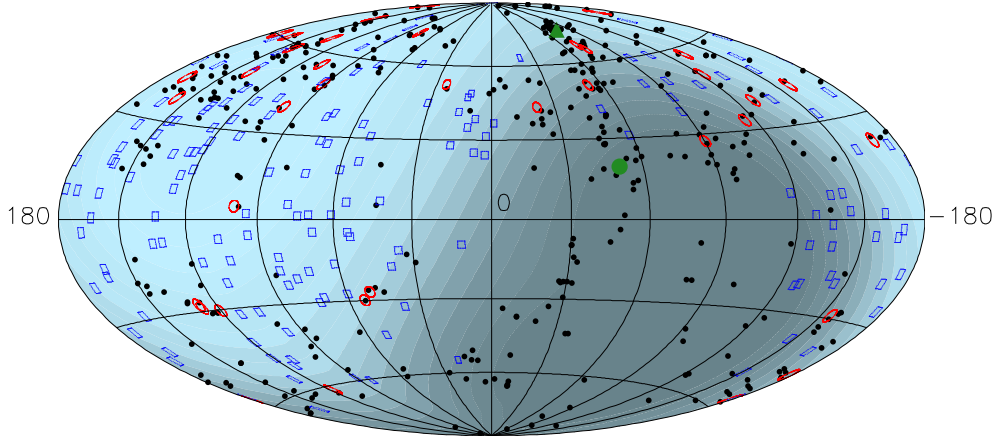}
  \end{center}
  \caption{A sky map of HiRes events above 16 EeV (blue squares and
    red boxes) and AGN with $z<0.016$ (black dots).  Correlated HiRes
    events are shown as red circles of radius $2.0^\circ$ while
    uncorrelated events are shown as blue squares.  The HiRes relative
    exposure is shown by the shading with the lightest shades being
    fully exposed and the darkes shade having no exposure.  There are
    11 shading bands for no exposure and then each 10\% of full
    exposure.}
  \label{corr-scan-point}
\end{figure}

We also scanned the parameter space ourselves, to find the best
correlation in our own data.  The scan was done according to the
Finley/Westerhoff method\cite{Finley-Westerhoff-2004-APP-21-359}, so
that we could correctly account for the scanning penalty and obtain an
unbiased random probability for our best correlation.  Our best fit
was with the parameters $E>16$ EeV, $\Delta\theta<2.0^\circ$ and
$z_{AGN}<0.016$.  We found 36 correlations among 198 events for
$P_{\rm min}=0.0018$.  However, when we look to see how often
simulated isotropic data sets of the same size have a smaller $P_{\rm
  min}$, we find that this happens in 24\% of these data sets.  Thus
the probability of getting the observed correlation somewhere within
our parameter space is an unconvincing 0.24.  The sky map of HiRes
events above 16 EeV, and AGN used as sources ($z<0.016$) along with
the HiRes exposure is shown in Figure~\ref{corr-scan-point}.

\section{Electron Neutrino Flux Limit}

By looking for upward going showers in a fluorescence detector, one
can measure or limit the flux of neutrinos, since all other particles
will be ranged out by the bulk of the Earth.  In addition, because of
the LPM effect\cite{Landau-Pomeranchuk-1953,Migdal-1956-PR-103-1811},
showers from electron neutrinos at high energies (above $10^{14}$ eV
in rock) develop very slowly, and thus can emerge from the ground with
significant particle numbers to be detectable starting from
interaction points very deep in the rock.  This gives a very large
target mass for electron neurtinos compared to what's available in the
atmosphere.  The Earth is largely opaque to neutrinos at these
energies, and the availale aperture come primarily from the limb of
the Earth, which would be evident as showers emerging from the ground
at large zenith angles.

A search was performed in the HiRes monocular data from the HiRes-II
site.  No such events were found, so a limit on the flux was set.
This result has now been accepted for
publication\cite{Abbasi-2008-arXiv-0803-0554}.

\begin{figure}
  \begin{center}
    \includegraphics[width=0.8\columnwidth]{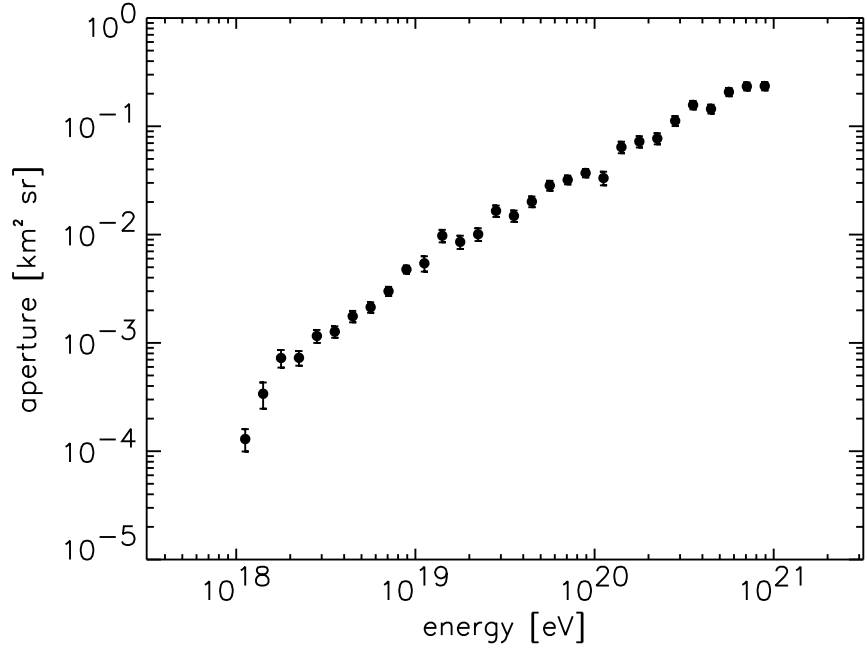}
  \end{center}
  \caption{An estimate of the aperture for upward going showers
    produced by electron neutrinos.  The aperture grows rapidly with
    energy due to the increase in available target mass due to the LPM
    effect.}
  \label{electron-neutrino-aperture}
\end{figure}

\begin{figure}
  \begin{center}
    \includegraphics[width=0.8\columnwidth]{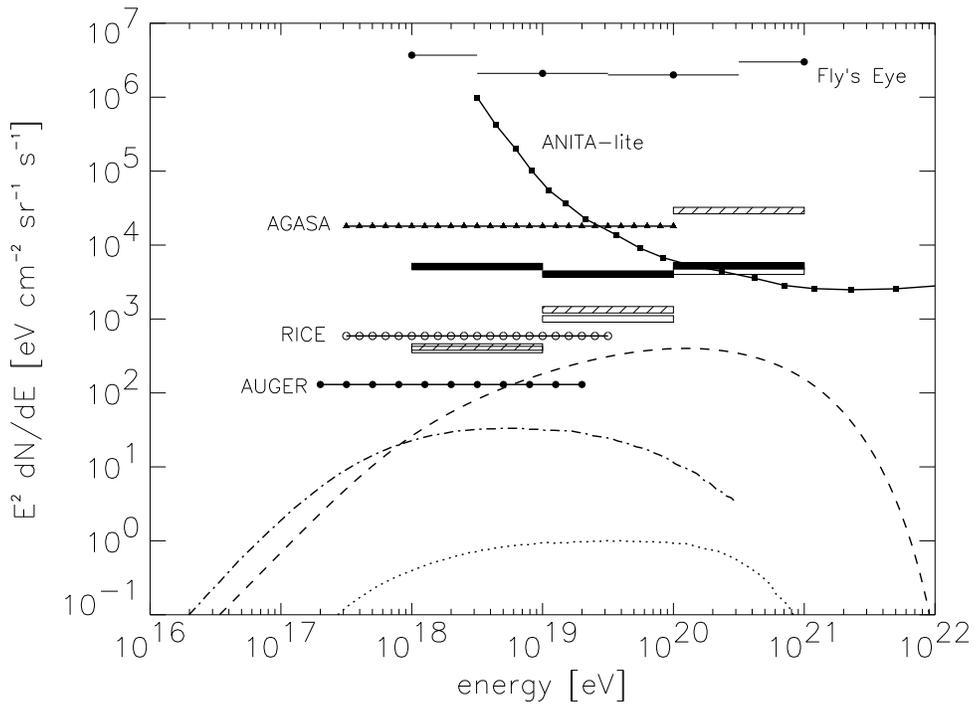}
  \end{center}
  \caption{The flux limit for electron neutrinos, integrated over
    decade bins in energy, is shown as the solid horizontal bars.  The
    HiRes tau-neutrino limit is shown as hatched bars, with the
    combination shown as open bars.  The current Auger limit and the
    older Fly's Eye electron neutrino limit are also shown.}
  \label{electron-neutrino-limits}
\end{figure}

To estimate the aperture of the HiRes detector in this mode, electron
neutrinos were thrown at random, and the probabilities of propagating
through the Earth and then interacting within a volume where the
shower would be visible to the detector were calculated.  To then
calculate the detector acceptance, showers were generated at rondom
along the neutrino path within the interaction region and the
fluorescence light propogated to the detector as for our spectrum
aperture calculation.  These simulated events were then subjected to
the same cuts used to define our search.  The calculated aperture
resulting from this method is shown in
Figure~\ref{electron-neutrino-aperture}.  Given the running time, and
the fact that we saw no events, we can convert this into a limit on
the electron neutrin flux.  This limit is shown in
Figure~\ref{electron-neutrino-limits} along with (and combined with) a
HiRes search for tau neutrinos.  The current
Auger\cite{2008-Abraham-PRL-100-211101} and older Fly's
Eye\cite{Baltrusaitis-1984-ApJ-281-L9} are also shown.  References for
the other shown limits and predictions can be found in the
preprint\cite{Abbasi-2008-arXiv-0803-0554} .

\section{Future Plans: Telescope Array}

Since neither the spectrum or the existence of source agree between
HiRes and Auger, or one might say between northern and southern
hemispheres, it is interesting to know what the future prospects are
for determining the source of these discrepancies: are they detector
related or due to viewing different parts of the universe.  To this
end the Auger collaboration hopes to build a detector in the North.

However, long before there is an Auger North detector, the Telescope
Array\cite{Fukushima-2007-ICRC-30} (TA) will answer many of these
questions.  TA is a hybrid detector (like Auger) combining the
fluorescence detectors of HiRes (in some cases equipment take directly
from the HiRes sites) with the scintillator based surface detectors of
AGASA (though not using the actual AGASA units).  It will have an
aperture of 1400 km$^2$ sr, and will thus match the total HiRes
exposure in about two years.  Most importantly it is already deployed
and is currently collecting data!

\section*{Acknowledgments}

This work was supported by US NSF grants PHY-9100221, PHY-9321949,
PHY-9322298, PHY-9904048, PHY-9974537, PHY-0073057, PHY-0098826,
PHY-0140688, PHY-0245428,\linebreak[4] PHY-0305516, PHY-0307098,
PHY-0649681, and PHY-0703893, and by the DOE grant FG03-92ER40732.  We
gratefully acknowledge the contributions from the technical staffs of
our home institutions. The cooperation of Colonels E.~Fischer,
G.~Harter and G.~Olsen, the US Army, and the Dugway Proving Ground
staff is greatly appreciated.

\section*{References}
\bibliography{DRB-Bibliography}
\end{document}